%% file: main.tex
\documentclass{article}
\usepackage{varwidth}

\input{acm-setting}
\graphicspath{{./figs/}}
\definecolor{NVgreen}{RGB}{118,185,0}
\definecolor{NVblack}{RGB}{0,0,0}
\definecolor{NVlgrey}{RGB}{205,205,205}
\definecolor{NVmgrey}{RGB}{140,140,140}
\definecolor{NVdgrey}{RGB}{94,94,94}

\definecolor{NVemerald}{RGB}{0,133,100}
\definecolor{NVamethyst}{RGB}{93,22,130}
\definecolor{NVintel}{RGB}{0,113,197}
\definecolor{NVgarnet}{RGB}{137,12,88}
\definecolor{NVfluorite}{RGB}{250,194,0}

\usepackage{authblk}

\begin{document}

\title{
   Generic Lithography Modeling with Dual-band \\ Optics-Inspired Neural Networks
}

\iftrue
{
\tiny
\author{Haoyu Yang$^1$, Zongyi Li$^{1,2}$, Kumara Sastry$^1$, Saumyadip Mukhopadhyay$^1$,
	Mark Kilgard$^1$,Anima Anandkumar$^{1,2}$,Brucek Khailany$^1$, Vivek Singh$^1$ and Haoxing Ren}             

\affil[1]{NVIDIA Corp.}
\affil[2]{California Institute of Technology}
}
\fi
\date{}
\maketitle

\begin{abstract}
	Lithography simulation is a critical step in VLSI design and optimization for manufacturability.
	Existing solutions for highly accurate lithography simulation with rigorous models are computationally expensive and slow, even when equipped with various approximation techniques.
	Recently, machine learning has provided alternative solutions for lithography simulation tasks such as coarse-grained edge placement error regression and complete contour prediction.
	However, the impact of these learning-based methods has been limited due to restrictive usage scenarios or low simulation accuracy.
	To tackle these concerns, we introduce an dual-band optics-inspired neural network design that considers the optical physics underlying lithography.
	To the best of our knowledge, our approach yields the first published via/metal layer contour simulation at 1$nm^2/$pixel resolution with any tile size.  
	Compared to previous machine learning based solutions, we demonstrate that our framework can be trained much faster and offers a significant improvement on efficiency and image quality with 20$\times$ smaller model size. 
	We also achieve 85$\times$ simulation speedup over traditional lithography simulator with $\sim 1\%$ accuracy loss.

\end{abstract}


\section{Introduction}
\label{sec:intro}
\begin{figure}[tb!]
	\centering
	\includegraphics[width=.7\textwidth]{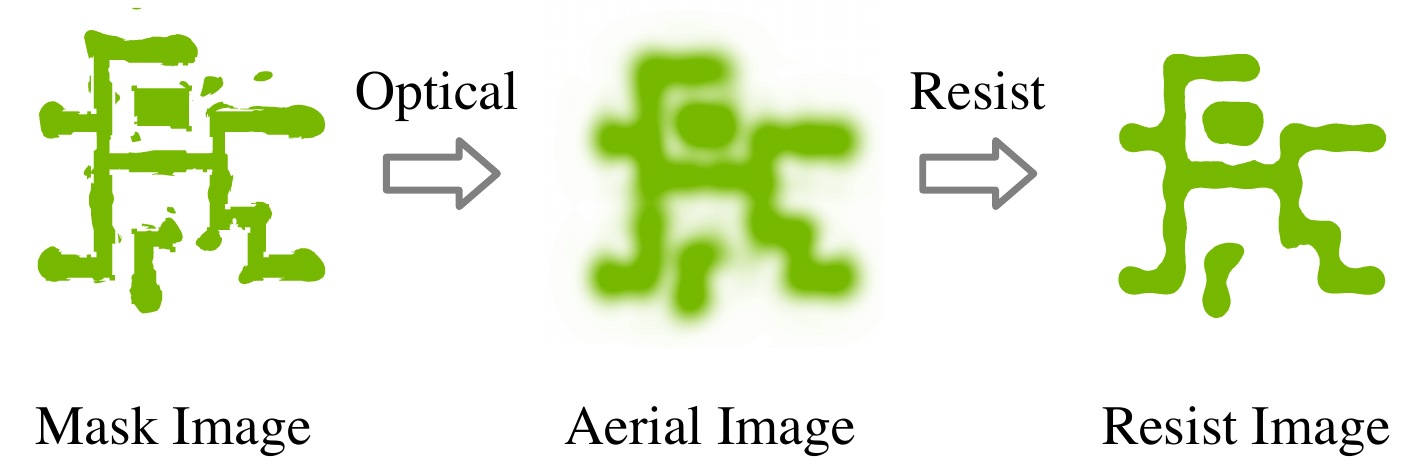}
	\caption{Lithography simulation with optical and resist model.}
	\vspace{-.1cm}
	\label{fig:intromodel}
\end{figure}

As the semiconductor industry continues to shrink geometries for each technology node, the requirements for efficient and high quality turnaround in manufacturability-aware layout design and optimization have increased \cite{ITRS}.
Lithography simulation is a critical procedure in design for manufacturability (DFM) flows as it significantly impacts the reliability and efficienty on mask optimization and layout printability estimation (see \Cref{fig:intromodel}) \cite{OPC-ICCAD2013-Banerjee,OPC-DAC2006-Yu,OPC-SPIE2003-Cobb}.
Existing solutions for highly accurate lithography simulation with rigorous models are computationally expensive and slow, even when equipped with various approximation techniques. \cite{DFM-B2011-Ma}.

Recently, machine learning (ML) has provided an alternative approach to costly physical simulation for DFM.  Some DFM applications include direct/indirect prediction of nominal edge placement error to guide mask optimization \cite{OPC-SPIE2015-Matsunawa,OPC-ASPDAC2019-Jiang} and direct lithography modeling with deep neural networks \cite{DFM-DAC2019-Ye,DFM-TCAD2019-Lin,OPC-ICCAD2020-DAMO,DFM-ISPD2020-Ye,DFM-SPIE2017-Shim,DFM-SPIE2019-Adam}. 
Matsunawa \textit{et al.~}\cite{OPC-SPIE2015-Matsunawa} build a Bayesian model to predict the  OPC edge movements.
Jiang \textit{et al.~}\cite{OPC-ASPDAC2019-Jiang} present a XG-Boost model to directly predict edge placement error at OPC control points and avoid computationally expensive lithography simulations.
Other recent efforts attempt to directly generate image contours with deep neural networks.
For example, \cite{DFM-TCAD2019-Lin,DFM-SPIE2017-Shim} propose to use ML techniques for resist modeling where \cite{DFM-TCAD2019-Lin} demonstrates the efficiency of transfer learning at the scenario of limited data access.
LithoGAN \cite{DFM-DAC2019-Ye} introduced conditional generative adversarial networks (CGAN) to predict resist image directly from mask patterns.
However, LithoGAN is only applicable to single contact prediction and requires additional CNN paths to generate coordinates that ensure the contour is aligned with the target mask.

TEMPO \cite{DFM-ISPD2020-Ye} developed another CGAN application that tackles the 3D-Mask challenge in EUV technology nodes to efficiently output an accurate aerial image from a thin mask image. 
A state-of-the-art deep learning-based lithography simulator (DLS) was proposed in DAMO \cite{OPC-ICCAD2020-DAMO} that exploits a nested-UNet generator and a carefully designed cost function.
The DLS was originally proposed to guide mask optimization tasks and is also the first work capable of predicting 4$\mu m^2$ contact layer contours at a 4$nm^2$/pixel resolution.
It should be noted that one major factor contributing to high quality contour generation is its input configuration, splitting each input mask into RGB channels containing SRAFs, OPC'ed contacts and design contacts respectively.
Such configurations, however, limit the usefulness of both DLS and LithoGAN, because a realistic lithography simulator must figure out which part of the mask should or should not be printed out. 
Also, significant improvements in resolution are still needed for accurate lithography modeling.

\textbf{Limitations of CNNs:}
We argue that a reason for these limitations in prior art is that {standard convolutional neural networks lack the ability to capture the necessary global information needed to accurately estimate lithography behavior}.
Although ML techniques such as dilated convolution \cite{dilate-conv} or deeper networks with more layers can increase the convolution receptive field and offer larger-scale feature understanding,
these methods may still not be effective enough.
Instead, a ML model should properly leverage both local high frequency and global low frequency components of a mask image for high lithography simulation accuracy.
Furthermore, spectral analysis has been studied on layout-related applications \cite{HSD-TCAD2019-Yang,HSD-JM3-2015-Shim}, showing that frequency domain representations can more powerfully reflect global layout characteristics.

\textbf{Fourier Neural Operator:}
Inspired by these findings, in this paper, we investigate the possibility of using a Fourier Neural Operator (FNO) \cite{PDE-ICLR2021-Li} for high-quality lithography modeling. 
The FNO concept was originally proposed as a neural network-based solver for partial differential equations (PDEs).  
It uses a series of Fourier layers that perform Fourier Transforms, learn mixing weights and convert back to spatial domain with Inverse Fourier Transforms on each feature map channel.
Although FNOs offer a great advantage for modeling physical equations, they are not directly applicable to lithography simulation tasks due to the large computation cost induced by multiple Fourier Transforms on extremely high-resolution images ($\sim 2000 \times 2000$).
Thus, we must adapt the approach to reduce computational costs while attaining the necessary information.

\textbf{Our Approach:}
To address these concerns, we propose an dual-band optics-inspired neural network (DOINN) for lithography modeling. 
The framework consists of two perception paths that handle low-frequency global information (global perception path) and high-frequency local information (local perception path) respectively.  
The global perception path consists of an optimized Fourier Unit that significantly reduces computing costs compared to the Fourier layer in FNO while incorporating the necessary global information to analyze lithography behavior. 
The local perception path uses convolutional layers to capture mask image details and compensate for the information loss in the global perception path. 
This is very close to the physics-informed neural network (PINN) used to solve PDEs in physics world \cite{PINN}.
However, instead of designing loss functions that reflect exact lithography behavior which is extremely challenging, 
we build a neural network architecture that allows approximated computing flow as in traditional lithography models, 
which adds an inductive bias to the model architecture \cite{PIML}.
We will show later that the optimized Fourier Unit resembles physical lithography equations, ensuring faster training convergence and contour generation quality.
To the best of our knowledge, this is the first publication of an optics-inspired neural network designed for high resolution metal/via layer lithography modeling at 1$nm^2$/pixel.
The major contributions of this paper include: 
(1) \textit{An optics-inspired dual band neural network design dedicated to learning lithography intensity and lithography contours.} 
(2) \textit{A large tile simulation scheme that allows our neural network to be trained on small tiles and simulate \textbf{ANY}-sized tiles}.

The remainder of the paper is organized as follows: \Cref{sec:prelim} introduces related terminologies and the problem formulation; 
\Cref{sec:alg} covers the details of the DOINN framework;
\Cref{sec:results} presents supporting experiments followed by the conclusion in \Cref{sec:conclu}.

\section{Preliminaries}
\label{sec:prelim}
In this section, we introduce our basic terminology and problem formulation. 
Throughout the paper, we use lowercase letters (e.g.~$x$) for scalars, bold lowercase letters for vectors (e.g.~$\vec{x}$), bold uppercase letters (e.g.~$\vec{X}$) for matrices, $\otimes$ for 2D convolution operations as in \texttt{PyTorch} and $\odot$ for element-wise production.
We use $\mathcal{F}$ and $\mathcal{F}^{-1}$ for Fourier and inverse Fourier transform, respectively.

\subsection{Lithography Approximation Model}
\begin{figure}[tb!]
	\centering 
	\includegraphics[width=.66\textwidth]{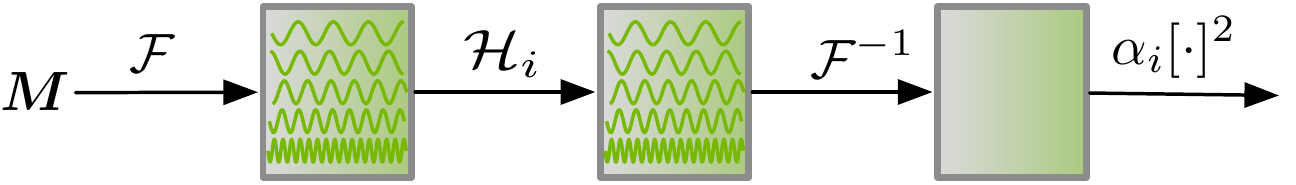}
	\caption{Lithography approximation model.}
	\label{fig:litho}
\end{figure}

The Hopkins diffraction model \cite{DFM-B2011-Ma} is well accepted in literature to represent lithography behavior.
However, computing the model is extremely time consuming.
To reduce the compute overhead, a singular value decomposition (SVD) approximation is typically adopted for lithography modeling.
The basic idea is to take the SVD of the coefficient matrix in the Hopkins model and formulate the lithography forward process as
\begin{align}
	\vec{I}(m,n)=\sum_{k=1}^{N^2} \alpha_k |\vec{h}_k(m,n) \otimes \vec{M}(m,n) |^2,
	\label{eq:svd-all}
\end{align}
where $\vec{h}_k$ terms are lithography kernels and $\alpha_k$ terms are the associated eigenvalues.
If we only keep the $l$ largest $\alpha_k$ values for faster calculation \cite{OPC-ICCAD2013-Banerjee,OPC-DAC2014-Gao}, \cref{eq:svd-all} becomes
\begin{align}
	\vec{I}(m,n)=\sum_{k=1}^{l} \alpha_k |\vec{h}_k(m,n) \otimes \vec{M}(m,n)|^2, l \ll N^2.
	\label{eq:svd-k}
\end{align}
The computing cost can be further reduced if we move to Fourier space as 
\begin{align}
	\vec{I}=\sum_{k=1}^{l} \alpha_k |\mathcal{F}^{-1} (\mathcal{F}(\vec{h}_k) \odot \mathcal{F}(\vec{M}))|^2,
	\label{eq:svd-f}
\end{align}
which is normally the equation used for forward simulation.
The intensity image calculation with respect to each lithography kernel can be abstracted as \Cref{fig:litho},
that includes a Fourier transform, a linear transformation described by lithography kernel in Fourier space and a nonlinear transformation.
For simplicity, we also apply a constant threshold resist model to obtain the final wafer contours.

\subsection{Problem Formulation}
It is straightforward to view lithography modeling as a pixel-level classification problem, therefore we adopt Mean Intersection Over Union (mIOU) and Mean Pixel Accuracy (mPA) as used in \cite{OPC-ICCAD2020-DAMO} to evaluate the performance of our framework.
mIOU and mPA are also widely used to measure semantic segmentation tasks.
\begin{mydefinition}[Mean Intersection Over Union (mIOU)]
	Given $k$ classes of predicted shapes $P_i$ and their ground truth $G_i$, $i=1,2,...,k$. The mIOU is defined as 
	\begin{align}
		\text{mIOU}(P,G)=\dfrac{1}{k} \sum_{i=1}^{k} \dfrac{P_i \cap G_i }{P_i \cup G_i}.
	\end{align}
\end{mydefinition}

\begin{mydefinition}[Mean Pixel Accuracy (mPA)]
	Given $k$ classes of predicted shapes $P_i$ and their ground truth $G_i$, $i=1,2,...,k$. The mPA is defined as 
	\begin{align}
		\text{mPA}(P,G)=\dfrac{1}{k} \sum_{i=1}^{k} \dfrac{P_i \cap G_i }{G_i}.
	\end{align}
\end{mydefinition}

It should be noted that in this paper we only focus on contour prediction, therefore only two classes (contour and background) are considered. With the above optimization targets, we now define the lithography modeling problem as follows:
\begin{myproblem}
	Given a set of mask images $\mathcal{M}_{tr}=\{\vec{M}_1,\vec{M}_2,...,\vec{M}_n\}$ and their corresponding wafer images $\mathcal{Z}_{tr}^\ast=\{\vec{Z}_1^\ast,\vec{Z}_2^\ast,...,\vec{Z}_n^\ast\}$,
	our target is to design and train a machine learning model $f(\cdot;\vec{W})$ such that for new designs $\mathcal{M}_{te}=\{\vec{M}_{n+1},\vec{M}_{n+2},...,\vec{M}_{n+k}\}$
	and $\mathcal{Z}_{te}^\ast=\{\vec{Z}_{n+1}^\ast,\vec{Z}_{n+2}^\ast,...,\vec{Z}_{n+k}^\ast\}$, mIOU$(f(\mathcal{M}_{te}), \mathcal{Z}_{te}^\ast)$ and mPA$(f(\mathcal{M}_{te}), \mathcal{Z}_{te}^\ast)$ can be maximized.
\end{myproblem}

\section{The Framework}
\label{sec:alg}
In this section, we cover the major components of DOINN, including a global perception path to capture layout semantic information and a local perception path for high frequency information understanding.
We also discuss the method for applying DOINN to large tile simulation.

\subsection{Dual-band Optics-Inspired Neural Network}
\subsubsection{Fourier Global Perception.}
\label{sec:gp}
The global perception path task is to capture mask semantic information in order to form resist contours.
Inspired by recent studies representing layout features in the frequency domain \cite{HSD-TCAD2019-Yang,HSD-SPIE2014-Shim} and the behavior of convolutional neural networks when extracting image high frequency and low frequency components \cite{DL-CVPR2020-Wang}, 
we propose an optimized Fourier unit that resembles a realistic optical model for mask layout global perception (GP).

\begin{figure}
	\centering
	\subfloat[Baseline FNO]{\includegraphics[width=.72\textwidth]{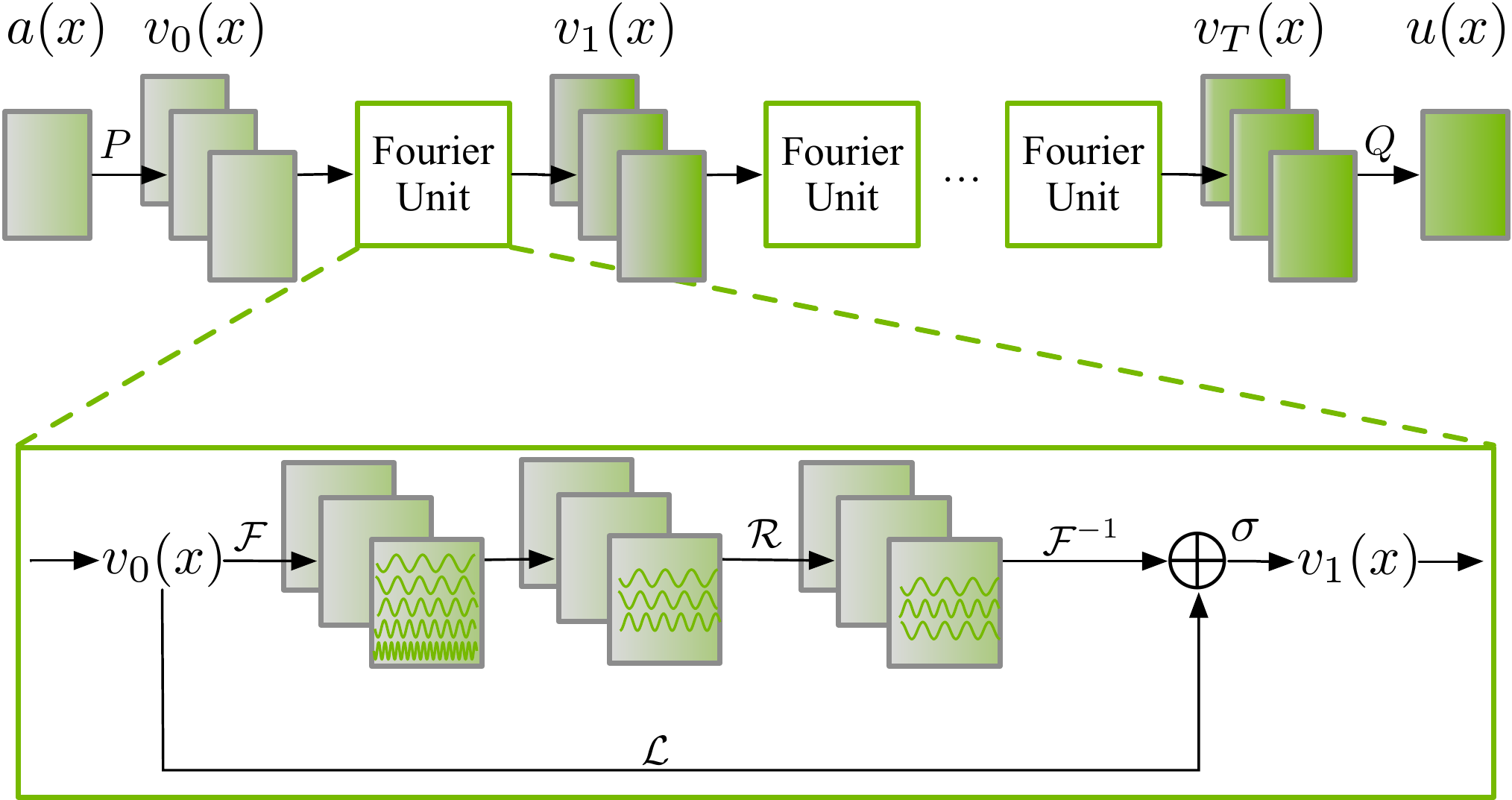} \label{fig:fno}}\\
	\subfloat[Reduced FNO]{\includegraphics[width=.72\textwidth]{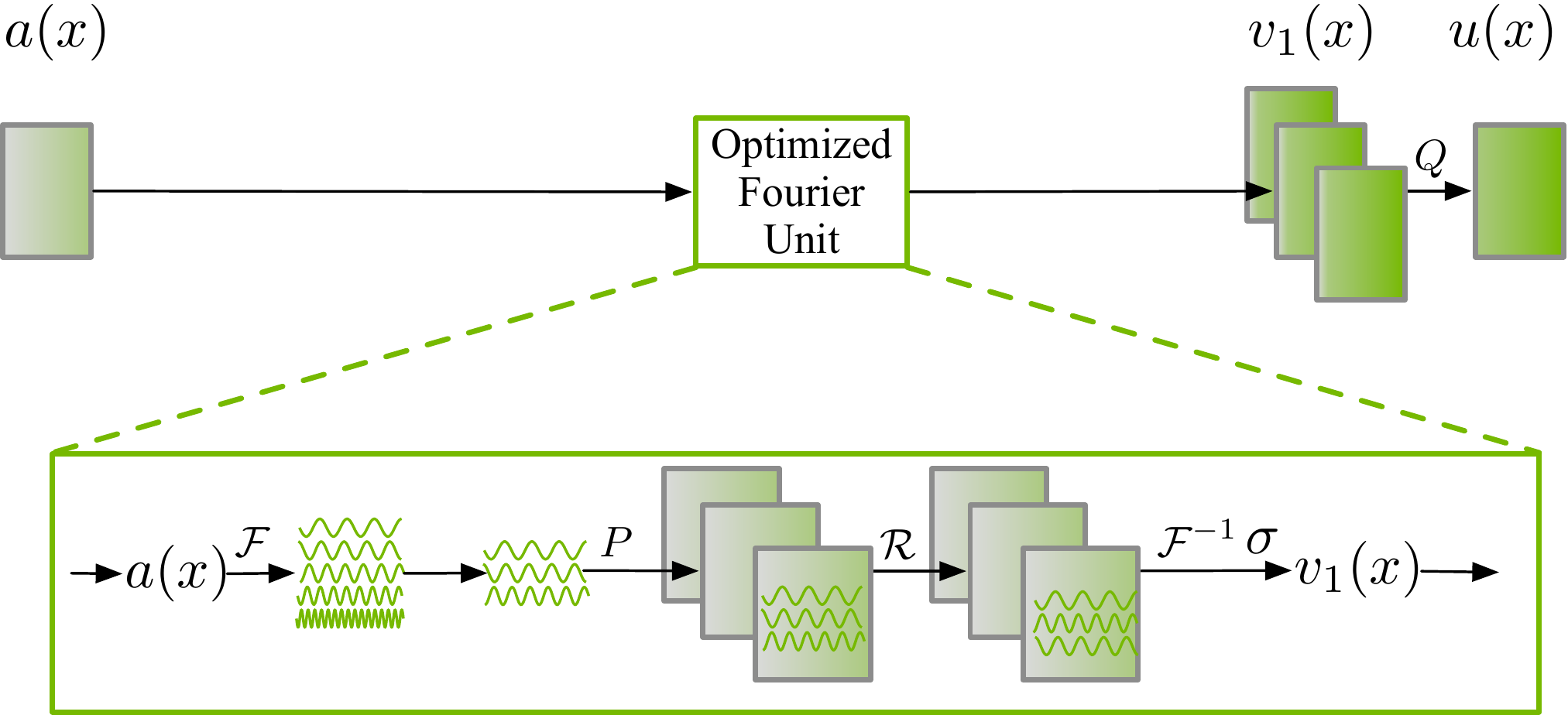} \label{fig:fno1l}}
	\caption{FNO architectures. (a) Baseline FNO built with stacked Fourier Units. (b) Reduced FNO with single optimized Fourier Units.}
	\label{fig:fnos}
\end{figure}

The Fourier unit was originally proposed as a Fourier Neural Operator (FNO) \cite{PDE-ICLR2021-Li} in a partial differential equation solver.  It tried to learn a parameterized mapping $\mathcal{G}$ between two finite dimension spaces from a set of observations such that $\mathcal{G}$ is close to the physical behavior.
\begin{align}
	\mathcal{G}: \mathcal{A} \times \Theta \rightarrow \mathcal{U},
\end{align}
where $\Theta$ is the parameter space. 
We define a baseline FNO solver as shown in \Cref{fig:fno},
where $a(x)$ and $u(x)$ are observed pairs from $\mathcal{A}$ and $\mathcal{U}$, $P$ is an operation that lifts $a(x)$ to a higher dimension and $Q$ will project the data back to the target dimension.
Inside each Fourier unit is a convolution operator defined on the Fourier space along with a bounded linear operation.
\begin{align}
	v_{t+1} (x) = \sigma ((\mathcal{L}v_t+\mathcal{F}^{-1} (\mathcal{R} \cdot (\mathcal{F} v_t))) (x)),
	\label{eq:fno}
\end{align}
where $\mathcal{L}$ and $\mathcal{R}$ are channel-wise linear transformations and $\sigma$ is some element-wise activation function.
Because we are dealing with image-to-image mapping in a lithography modeling problem, \cref{eq:fno} will work in a discrete manner.
Let the input of each Fourier Unit to be a 3-D tensor $\vec{V}_t \in \mathbb{R}^{C \times H \times W}$, then \cref{eq:fno} becomes
\begin{align}
	\vec{V}_{t+1} = \sigma(\vec{V}_{t,\mathcal{L}} + \vec{V}_{t,\mathcal{F}}),
\end{align}
where
\begin{align}
	\vec{V}_{t,\mathcal{L}} &= \vec{V}_t \otimes \vec{W}_{t,\mathcal{L}}, \vec{W}_{t,\mathcal{L}} \in \mathbb{R}^{1 \times C  \times 1 \times 1},\\
	\vec{V}_{t,\mathcal{F}} &= \mathcal{F}^{-1}(\mathcal{F}(\vec{V}_t)_{k-\text{truncated}} \cdot \vec{W}_\mathcal{R}), \vec{W}_\mathcal{R}  \in \mathbb{C}^{C \times C  \times H \times W}.
	\label{eq:fu}
\end{align}
Note that all the Fourier transforms and inverse Fourier transforms are conducted on the last two dimensions of each tensor only and
$\mathcal{F}(\vec{V}_t)_{k-\text{truncated}}$ is obtained by only keeping the $k$ lowest frequency components of $\mathcal{F}(\vec{V}_t)$ and discarding the remaining entries (by setting to zero).
It can be observed that there are two major concerns for the baseline FNO being an efficient lithogrpahy simulator:
\begin{itemize}
	\item Stacked Fourier Units are not compliant with the real physical model that comes with an single Fourier Transform on mask and inverse Fourier Transforms to get resist light intensity.
	\item Multiple Fourier and inverse Fourier transforms pose large computational overheads when processing large layout tiles.
\end{itemize}

To address these concerns, we propose a reduced FNO architecture with a single Fourier Unit (see \Cref{fig:fno1l}).
Since Fourier transforms are applied channel-wise and preserve linearity, we develop an optimized Fourier Unit that performs Fourier transforms prior to the channel-lifting operation $P$.
Thus the computation inside the Fourier Unit becomes 
\begin{align}
	\vec{V} = \sigma(\mathcal{F}^{-1}(\mathcal{F}(\vec{A})_{k-\text{truncated}} \otimes \vec{W}_P \cdot \vec{W}_\mathcal{R})),
	\label{eq:ofu}
\end{align}
where $\vec{A} \in \mathbb{R}^{1 \times H \times W}$ is the input image, and $\vec{W}_P \in \mathbb{C}^{1 \times C \times 1 \times 1}$ is the convolution kernel that performs channel-lifting.
Since the new design only consists of one Fourier Unit, the bypass link $\mathcal{L}$ becomes less necessary, so we discard it to simplify the network.  With these optimizations, \textit{\cref{eq:ofu} attains the same expressive power as \cref{eq:fu} while saving a significant amount of calculation}.
In \cref{eq:ofu,eq:fu}, convolution operations cost $\mathcal{O}(C^2 HW)$ since typically $C \ll H,W$.
Assuming both $\mathcal{F}$ and $\mathcal{F}^{-1}$ are implemented with Fast Fourier Transforms (FFT) and complete in $\mathcal{O}(HW\log(HW))$,
FFTs would dominate the computation. 
Given that FFTs are performed channel-wise, our new architecture defined in \cref{eq:ofu} can save $\sim50\%$ of runtime by saving $C$ FFT operations.
Also, \cref{eq:ofu} is specifically designed with Fourier Transform, linear transformation, Inverse Fourier Transform and nonlinear operations to match the simulation pipeline as in \Cref{fig:litho}, but with significantly less computing cost.

It should be noted that we have discarded most high frequency components in the Fourier Unit, so it is no longer necessary to keep a high resolution input $\vec{V}$.
We will show later how that can further help design an efficient lithography simulator with a down-sampled input.

\subsubsection{Convolutional Local Perception.}
\label{sec:lp}
Although the reduced FNO is designed to capture the semantic information of mask images, it alone is not sufficient for precise lithography modeling.
The primary reason is that the Fourier Unit drops most high frequency components of a mask image which contribute to detail formation and are equally important as semantic information.
A recent study \cite{DL-CVPR2020-Wang} showed that convolutional neural networks are capable of understanding high frequency components in images,
so we therefore introduce a convolutional local perception (LP) path to compensate for the detail loss introduced with the Fourier Unit optimization.
Here, we adopt stacked VGG \cite{VGG} blocks for image local perception to generate learned feature maps carrying high frequency mask content.

\subsubsection{DOINN Architecture.}
\label{sec:DOINN}
\begin{figure*}
	\centering
	\includegraphics[width=1\textwidth]{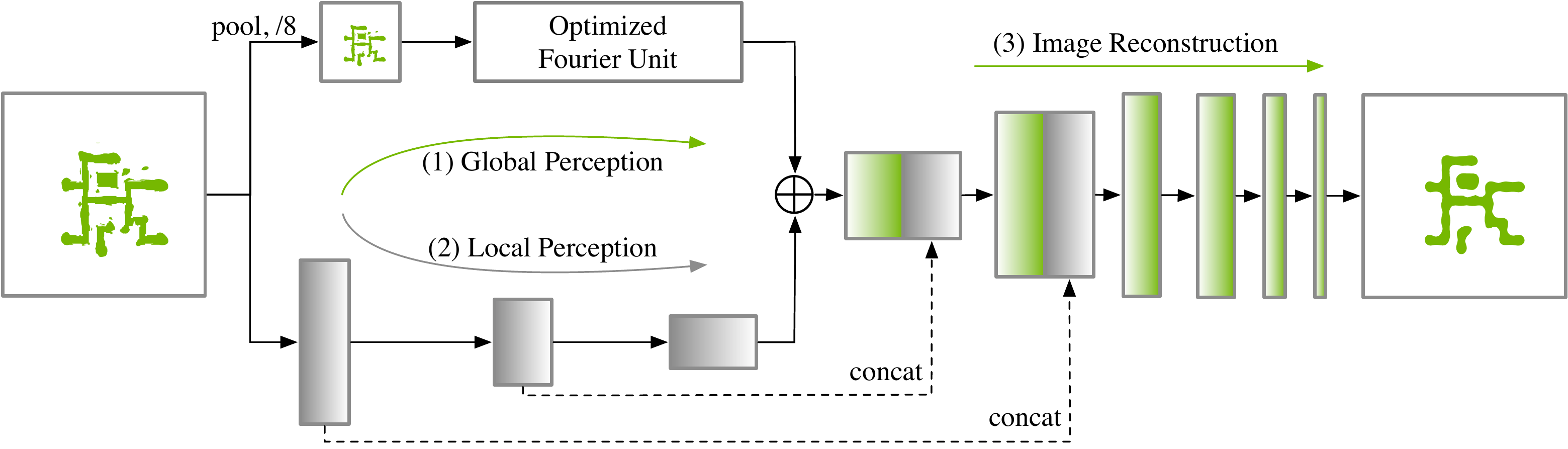}
	\caption{The overall contour prediction pipeline of the DOINN.}
	\label{fig:DOINN}
\end{figure*}

The DOINN components described previously capture mask global and local information. 
The remaining task is to collect global and local perception features and use them to reconstruct resist contours.  To achieve this, we propose a convolution-based image reconstruction (IR) path that takes the concatenated GP and LP feature maps and generates the desired resist image.
Essentially, IR consists of a series of transposed convolution layers to rescale low level feature maps back to the original mask size followed by single-strided convolution layers for contour refinement.
The three paths of GP, LP and IR together form our dual-band neural network, as shown in \Cref{fig:DOINN}.
Because the Fourier Unit operates in a mechanism that resembles the physical lithography model as in \cref{eq:svd-f}, we claim the DOINN to be optics-inspired. 
In summary, the DOINN has the following advantages: 
(1) The optimized Fourier Unit performs an efficient global perception on input mask images from frequency domain analysis.
(2) Convolutions in the LP path compensate for the high frequency information loss in the GP path to produce high quality resist contour reconstructions. 
(3)The optics-inspired property of DOINN 
makes it possible to create highly accurate lithography modeling results with a $10 \times$ smaller model size than prior art.

\vspace{-.1cm}
\subsection{Large Tile Simulation Scheme}
\label{sec:fullchip}
Because the proposed DOINN is differentiable, it is straightforward to train it with supervised learning and evaluate with fix-sized layout clips, as has been commonly been presented in literature.
However, to demonstrate utility on practical tasks, we target a more challenging task:  large tile simulation.
Although our DOINN configuration in \Cref{fig:DOINN} allows for any-sized input, the generated resist image quality will be seriously affected if the input tile size is much larger than the training tile size.
This is because the weights in the Fourier Unit are trained for the $k$-lowest frequency components of a smaller tile and will result in significant information loss if applied on scaled inputs.

\begin{figure}
	\centering
	\includegraphics[width=.72\textwidth]{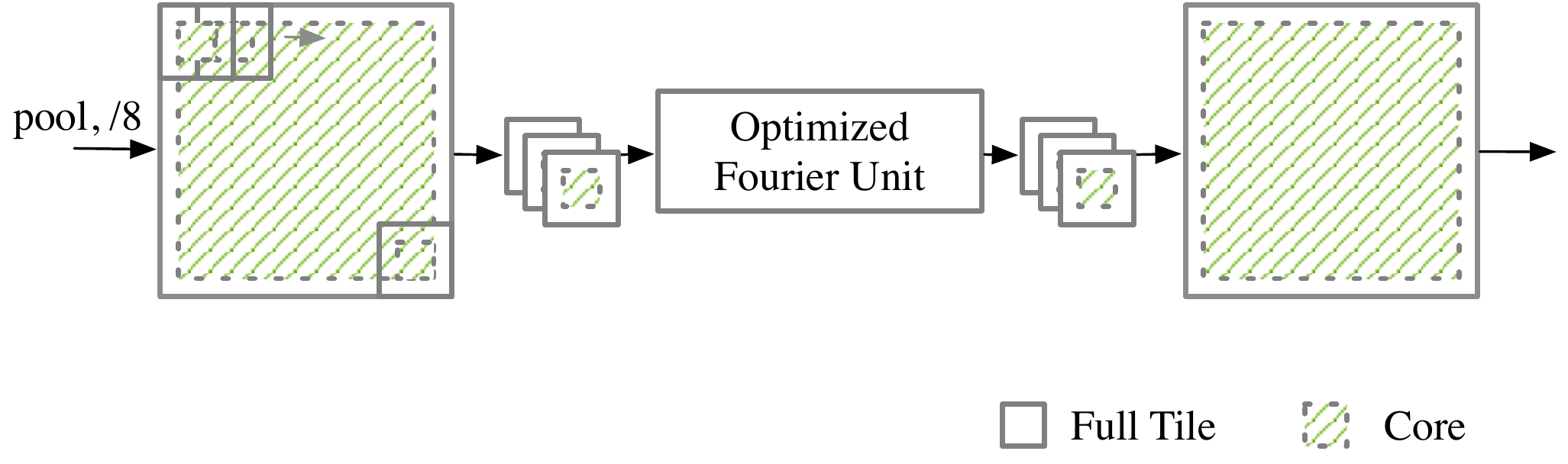}
	\caption{Large tile global perception.}
	\label{fig:DOINN-full}
\end{figure}

Intuitively, we can cut the clips into tiles that have the same size as the clips used for DOINN training.
This is, however, not as simple as it looks due to the concept of \textit{optical diameter}.
Lithography physics tells us that the light intensity on the resist stage can be viewed as a superposition of multiple orders of diffraction patterns.  As a result,
the light intensity at a certain wafer location is determined by a large area of surrounding mask patterns.
The size of this area is usually measured by its diameter, i.e.~the optical diameter \cite{DFM-B2011-Ma,DFM-SPIE2017-Tabery}.
For this reason, simply cutting the large tile into small clips is not feasible as we must account for shapes near the tile boundaries.
To resolve this issue, we propose a large tile global perception path as illustrated in \Cref{fig:DOINN-full}.
The green region at the center of each tile is the core mask region that we want to simulate.  The shapes outside the core region will be ignored as they are too close to the boundaries.
The basic idea is to cut the large tile into small clips which have the same size as is used for training, fed into the Fourier Unit in batches.
It should be noted that small tiles are half overlapped with each other such that the core regions in these tiles will exactly cover the core region of a large tile input.
The output feature maps will then be concatenated back to the large tile size.
Nothing needs to be done with the remaining convolution and transposed convolution layers due to the shared feature maps in the calculation pipeline.

This large tile global perception can be mathematically described in the following manner.
Suppose the DOINN is trained with $H \times W$ mask-contour pairs and the optical diameter is $d$.
The global perception $\mathcal{G}: \mathbb{R}^{H \times W} \rightarrow \mathbb{R}^{C \times \frac{H}{8} \times\frac{W}{8}}$ can be expressed as
\begin{align}
	\vec{F}_\text{gp} = \mathcal{G} (\vec{M}; \vec{W}_\mathcal{P}, \vec{W}_\mathcal{R}),
\end{align}
where $\vec{F}_\text{gp} \in \mathbb{R}^{C\times \frac{H}{8} \times \frac{W}{8}}$ is the feature map output of GP path, $\vec{M} \in \mathbb{R}^{H \times W}$ denotes the after-pooling mask and $\vec{W}_\mathcal{P}, \vec{W}_\mathcal{R}$ are trained parameters of the Fourier Unit.
Let $\mathcal{G}_\text{s} : \mathbb{R}^{sH \times sW} \rightarrow \mathbb{R}^{C \times \frac{sH}{8} \times \frac{sW}{8}}$ be the global perception path which processes a mask $\vec{M}_\text{s} \in \mathbb{R}^{sH \times sW}$ that is $s \times$ larger than tiles used for training.
Then each entry of output feature map $\vec{F}_\text{s,gp}$ is defined as
\begin{align}
	&\vec{F}_\text{s,gp}[:,i,j] = \mathcal{G}_\text{s} (\vec{M}_\text{s}; \vec{W}_\mathcal{P}, \vec{W}_\mathcal{R})[:,i,j] \nonumber \\
	=&\mathcal{G} (\vec{M}_\text{s} [\frac{mH}{2}:\frac{(m+2)H}{2},\frac{nW}{2}:\frac{(n+2)W}{2}]; \vec{W}_\mathcal{P}, \vec{W}_\mathcal{R})[:,p,q],
\end{align}
where $\frac{d}{2} \le i < sH-\frac{d}{2}$, $\frac{d}{2} \le j < sW-\frac{d}{2}$ and 
\begin{align}
	m &= \floor{\frac{2(i-\frac{d}{2})}{H}}, n = \floor{\frac{2(j-\frac{d}{2})}{W}}, \nonumber \\
	p &= ((i - \frac{d}{2}) \mod \frac{H}{2}) + \frac{d}{2}, \nonumber \\
	q &= ((j - \frac{d}{2}) \mod \frac{W}{2}) + \frac{d}{2}.
\end{align}
It should be noted that \textit{this large tile simulation scheme allows our framework to be trained with small tiles and evaluated on any sized chip input without performance degradation}.

\section{Experiments}
\label{sec:results}

\subsection{The Dataset}
We evaluate our framework on both metal layer and via layer designs to demonstrate accuracy and generality.
Details are listed in \Cref{tab:data}.
\subsubsection{ISPD-2019.} 
This design comes from the ISPD 2019 initial detailed routing contest \cite{ISPD2019C}.  The layout is synthesized with commercial placement and routing tools.
Nominal resist contours are obtained by feeding the via layer into OPC and simulation flows.
The training set comprises 10K 4$\mu m^2$ tiles generated using an open source layout generator following the same design rules as designs in \cite{ISPD2019C}.
For testing, via layers in \cite{ISPD2019C} are also cut into 4$\mu m^2$ tiles with total 11K instances, which are converted to $2000 \times 2000$ (ISPD-2019 (H)) and $1000 \times 1000$ (ISPD-2019 (L)) images.
In the experiment of large tile simulation, we take the ten most dense 64$\mu m^2$ tiles from \cite{ISPD2019C} and convert them to $8000 \times 8000$ images.

\subsubsection{ICCAD-2013.}
This design contains ten 4$\mu m^2$ tiles from the ICCAD 2013 CAD Contest \cite{OPC-ICCAD2013-Banerjee}. 
The training set contains $\sim$4K 4$\mu m^2$ tiles which are generated in a manner similar to that used for training GAN-OPC \cite{OPC-TCAD2020-Yang}. 
The OPC'ed masks are generated using MOSAIC \cite{OPC-DAC2014-Gao} and contours are simulated with \texttt{Lithosim} from \cite{OPC-ICCAD2013-Banerjee}.

\subsubsection{N14.}
This dataset contains a single via layer in a 14$nm$ technology node, cut into 1.6K $4\mu m^2$ clips.
Each clip is converted to a $2000\times2000$ image.
We take out the 100 densest clips for evaluation and use the remaining clips for training.


\begin{table}[tb!]
	\centering
	\caption{Details of the Dataset.}
	\label{tab:data}
	\setlength{\tabcolsep}{3pt}
	\renewcommand{\arraystretch}{1}
	\begin{tabular}{c|cccc}
		\toprule
		Dataset  & Train & Test  & Tile Size &Litho Engine \\ \midrule
		ICCAD-2013    & 4875  & 10    & 4$\mu m^2$ & Lithosim \cite{OPC-ICCAD2013-Banerjee}  \\
		ISPD-2019      & 10300 & 11641 & 4$\mu m^2$ & Calibre \cite{Calibre}  \\
		ISPD-2019-LT    &   -   & 10    & 64$\mu m^2$  & Calibre \cite{Calibre}\\ 
		N14 &1630 &137 &4$\mu m^2$   & - \\ \bottomrule
	\end{tabular}
\end{table}


\subsection{Result Comparison with State-of-the-Art}
We first compare our framework with the state-of-the-art ML-based lithography simulator DAMO-DLS \cite{OPC-ICCAD2020-DAMO} and a popular ML model with a UNet structure \cite{UNet}.
The results are listed in \Cref{tab:result}, where 
columns ``mPA (\%)'' and ``mIOU (\%)'' represent the mean pixel accuracy and mean intersection over union as defined in \Cref{sec:prelim}.  
Column ``UNet'' corresponds to the result of the UNet model \cite{UNet},
column ``DAMO-DLS'' lists the results of the deep lithography simulator in \cite{OPC-ICCAD2020-DAMO},
and column ``Ours'' lists the results of the DOINN framework proposed in this paper.
In rows ``ISPD-2019 (L)'' and ``ICCAD-2013 (L)'', we show the results when 4$\mu m^2$ tiles are converted to $1000 \times 1000$ images. 
Rows ``ISPD-2019 (H)'' and ``ICCAD-2013 (H)'' are the corresponding  results on high resolution images with 4$\mu m^2$ tiles rendered in $2000 \times 2000$.
Row ``N14'' corresponds to the results on 14$nm$ high-density vias.
We do not include a result comparison with LithoGAN \cite{DFM-DAC2019-Ye} because the model in \cite{DFM-DAC2019-Ye} only supports the output of a single via shape in a small area.
It should also be noted that ``DAMO-DLS \cite{OPC-ICCAD2020-DAMO}'' only supports $1000 \times 1000$ inputs. 

As can be seen in the result table, our framework outperforms state-of-the-art ML models in terms of mPA and mIOU on both metal and via layers and on both low and high resolution images.
For the via cases, all three models exhibit similar results with DOINN being slightly better because via designs are regular and relatively easy to learn.
For metal shapes, results show a larger performance gap from our DOINN framework to other models: 2\% better mIOU than UNet and 1\% better than DAMO-DLS.
When it comes to designs in advanced technology nodes, our method  exhibits even more advantages over previous solutions, achieving 4\% higher mPA and 5\% higher mIOU over \cite{UNet}. 

\begin{figure}
	\centering
	\input{time.tex}
	\caption{Runtime comparison with state-of-the-art.}
	\label{fig:runtime}
\end{figure}
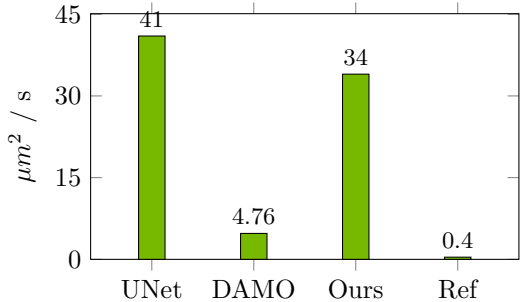

\Cref{fig:runtime} shows the runtime comparison of the three models in term of throughput ($\mu m^2 / s$). We can see that the DOINN framework not only converges much faster but also exhibits much faster inference time than DAMO \cite{OPC-ICCAD2020-DAMO}.
This can be attributed to the 20$\times$ smaller model size of DOINN (1.3M parameters) when compared to DAMO-DLS (18M parameters).
``Ref'' corresponds to the average throughput of traditional lithograph engines used to obtain golden simulation contours, 
where we can observe >80$\times$ speedup of our framework compared to traditional lithography engines with $\sim 1\%$ accuracy loss.

\begin{table}[tb!]
	\centering
	\vspace{.3cm}
	\caption{Result Comparison with State-of-the-Art.}
	\label{tab:result}
	\setlength{\tabcolsep}{2pt}
	\renewcommand{\arraystretch}{1}
	\begin{tabular}{c|cc|cc|cc}
		\toprule
		\multirow{3}{*}{Benchmark} & \multicolumn{2}{c|}{UNet \cite{UNet}}  & \multicolumn{2}{c|}{DAMO-DLS \cite{OPC-ICCAD2020-DAMO}} & \multicolumn{2}{c}{Ours} \\
		& mPA   & mIOU     & mPA      & mIOU   & mPA   & mIOU   \\ 
		& (\%)   & (\%)     & (\%)      & (\%)   & (\%)   & (\%)   \\ \midrule
		ISPD-2019 (L)             &   99.40    &    98.03   &   99.25       &    98.11     &   \textbf{99.43}    &  \textbf{98.27}     \\
		ISPD-2019 (H)             &   99.08    &    97.97   &   -           &     -        &   \textbf{99.21}    &  \textbf{98.45}      \\
		ICCAD-2013 (L)            &   97.30    &    95.38   &   98.94       &    96.97     &   \textbf{98.98}    &  \textbf{97.79}      \\
		ICCAD-2013 (H)            &   95.16    &    93.04   &   -           &    -         &   \textbf{99.12}    &  \textbf{97.77}      \\
		N14 & 94.39 & 91.64 & - & -  & \textbf{98.68}  & \textbf{96.49} \\ \bottomrule
	\end{tabular}
\end{table}

\begin{table}[tb!]
	\centering
	\caption{Ablation Study.}
	\label{tab:ablation}
	\setlength{\tabcolsep}{6pt}
	\renewcommand{\arraystretch}{1}
	\begin{tabular}{c|cccc|cc}
		\toprule
		\multirow{2}{*}{ID} &	\multicolumn{4}{c|}{Technique}                 & \multicolumn{2}{c}{ICCAD-2013 (L)} \\
		&GP & IR & LP & ByPass & mPA (\%)         & mIOU (\%)        \\ \midrule
		1&\checkmark   &              &                  &        &     97.50        &      96.09       \\
		2&\checkmark   & \checkmark       &             &         &     98.40        &      97.20       \\
		3&\checkmark   & \checkmark     & \checkmark     &        &     98.79  &            97.60 \\
		4&\checkmark  & \checkmark     & \checkmark & \checkmark  &      \textbf{98.98}       &       \textbf{97.79}      \\ \bottomrule
	\end{tabular}
\end{table}



\subsection{Ablation Study}
We also conduct experiments to study the effectiveness of each component technique of the DOINN.
Benchmark ``ICCAD-2013 (L)'' is applied as an example for simplicity.
As shown in \Cref{tab:ablation}, there are four groups of studies that gradually enable all designed features in DOINN. 
Group ID ``1'' corresponds to the experiment using the Fourier Unit only.
Group ID ``2'' adds four single stride convolution layers to further refine the contour.
Group ID ``3'' further includes the convolutional local perception path with exactly the same structure as in \Cref{fig:DOINN} except with two concatenate links.
Finally, group ID ``4'' corresponds to the full version of DOINN.
From the result table we observe that each component in our DOINN structure improves the resist prediction quality.
In particular, the local perception can significantly increase the prediction performance at the cost of a very small number of additional parameters.

\iftrue
\vspace{-.1cm}
\subsection{Learning Global and Local Perception}
\begin{figure}[tb!]
	\centering
	\subfloat[Global Perception]{\includegraphics[width=.38\textwidth]{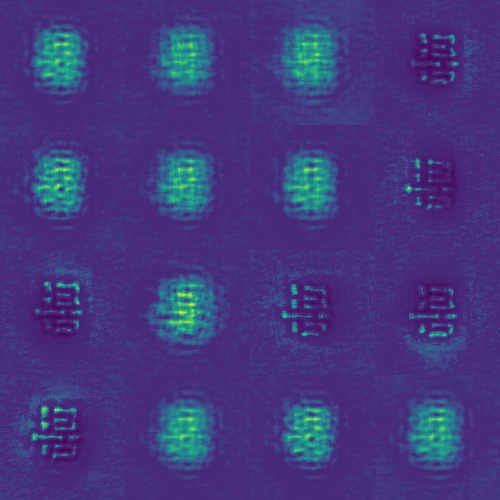} \label{fig:gpp}}
	\subfloat[Local Perception]{\includegraphics[width=.38\textwidth]{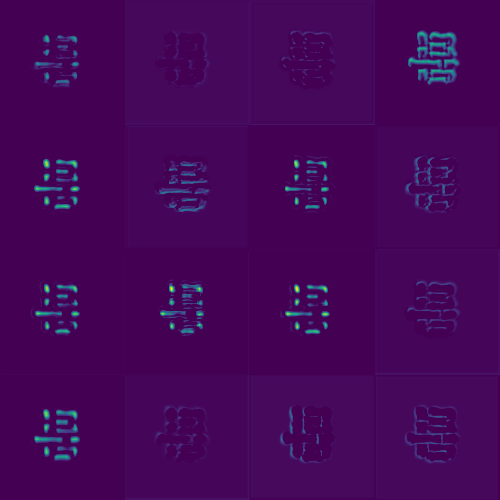} \label{fig:lpp}}
	\caption{Feature map visualization of (a) GP and (b) LP paths.}
	\vspace{6pt}
	\label{fig:DOINN-feature}
\end{figure}

To further understand the behavior of DOINN, we visualize the feature maps of GP and LP paths.
\Cref{fig:gpp} shows the output of the Fourier Unit. We observe that each channel of GP output is able to capture the intensity map that is close to the aerial image.
On the other hand, \Cref{fig:lpp} includes feature maps learned through the convolutional LP path, mainly focused on shape edges and contour details.
We can see both perception paths trying to understand the lithography procedure globally and locally, consistent with our expectation for DOINN.
\fi

\subsection{Sensitivity to Subtle Perturbations}
\begin{figure}
	\centering 
	\input{24opc.tex}
	\caption{Lithography modeling performance on subtle perturbations.}
	\label{fig:24opc}
\end{figure}
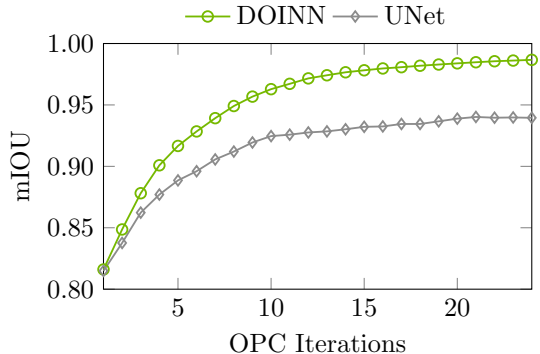

One key motivation of developing fast lithography modeling framework is to perform layout printability check prior manufacturing or on the OPC runtime.
This requires a lithography simulator to be able to identify subtle perturbations on the mask.
Therefore, we pick up mask images of a metal layer design at 24 different OPC iterations and feed them for resist image prediction. 
The mIOU for DOINN and UNet are depicted in \Cref{fig:24opc}.
Because both networks are trained towards the OPC'ed mask and resist image pairs, we can observe weak behavior at early OPC iterations when mask images are close to design images.
Thanks to the inductive bias introduced from the Optimized Fourier Unit, DOINN still exhibits significant advantage over CNN-based networks.

\subsection{Large Tile  Simulation}
Finally, we evaluate the ability of the proposed DOINN to handle large tile simulations.
In this experiment, we still train the DOINN with $2000 \times 2000$ tiles covering a 4$\mu m^2$ layout area. 
For the sake of simplicity and memory consumption, we do not feed an entire chip into the DOINN.
Instead, we verify the large tile simulation scheme in \Cref{sec:fullchip} with ten 64$\mu m^2$ tiles with high via density.
\Cref{tab:fs} lists the simulation results on large test tiles.
``DOINN'' corresponds to the results of feeding each tile into the default DOINN pipeline.  ``DOINN-LT'' implements the proposed large tile simulation scheme.
We can clearly observe that the default DOINN simulation pipeline suffers a large accuracy drop when fed with tiles larger than those used for training due to low frequency information loss.
This can be reflected as noise in predicted resist contours, as depicted in \Cref{fig:8k}.
However, with the help of the proposed large tile simulation scheme, we are able to attain the ability to generate high quality resist contours.

\iftrue
\begin{figure}[tb!]
	\centering
	\subfloat[Mask]{\includegraphics[width=.24\textwidth]{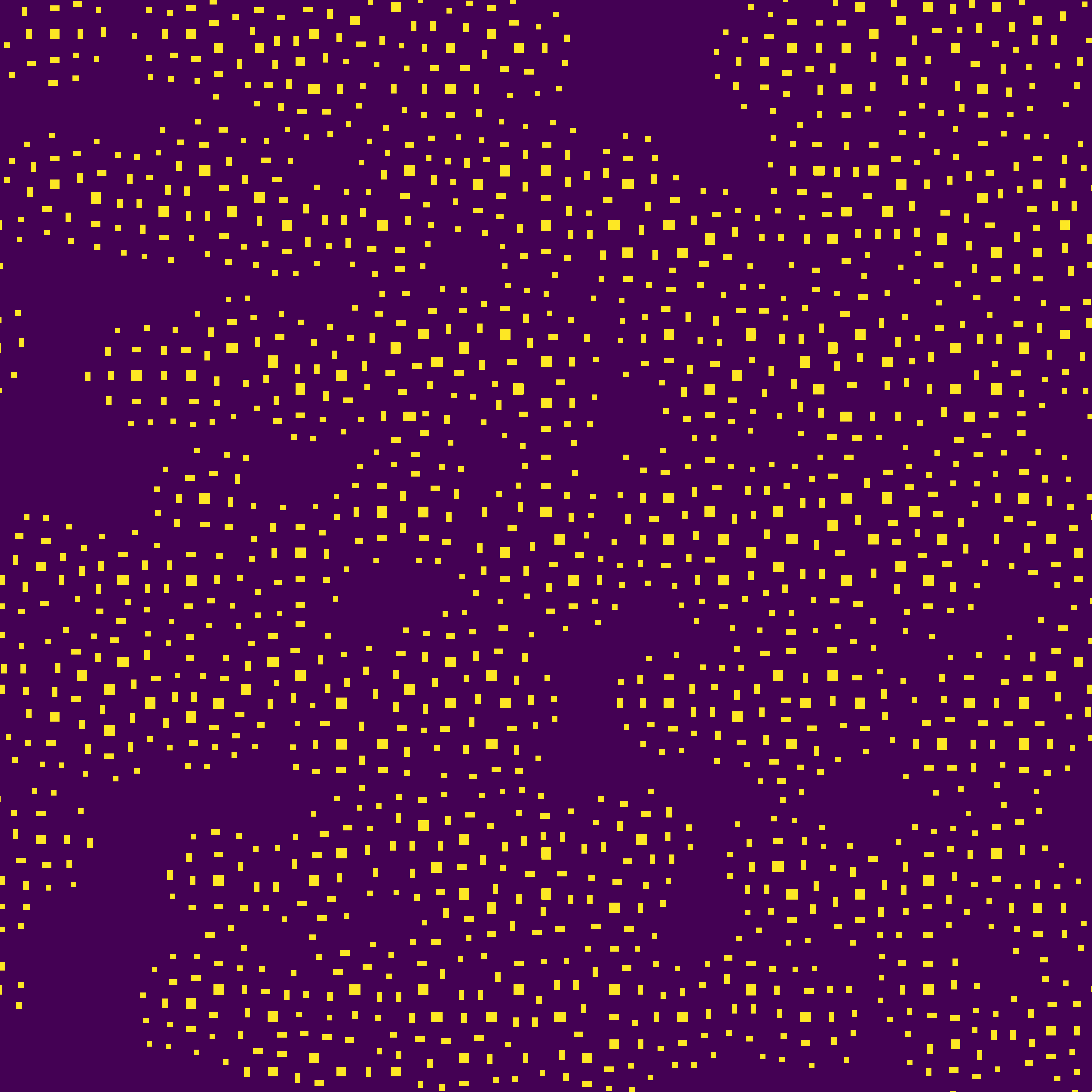}} \hspace{0.1cm}
	\subfloat[DOINN]{\includegraphics[width=.24\textwidth]{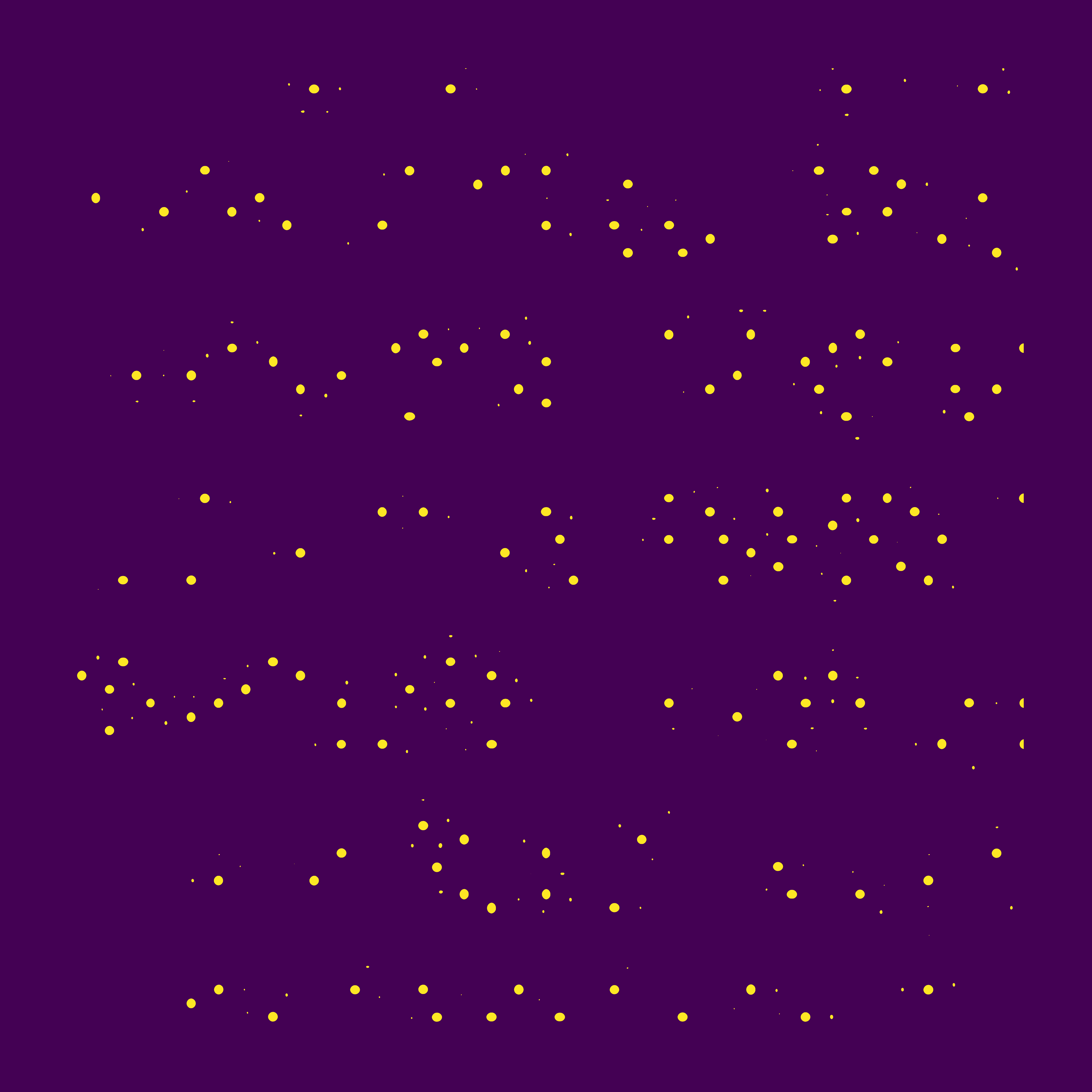}} \hspace{0.1cm}
	\subfloat[DOINN-FS]{\includegraphics[width=.24\textwidth]{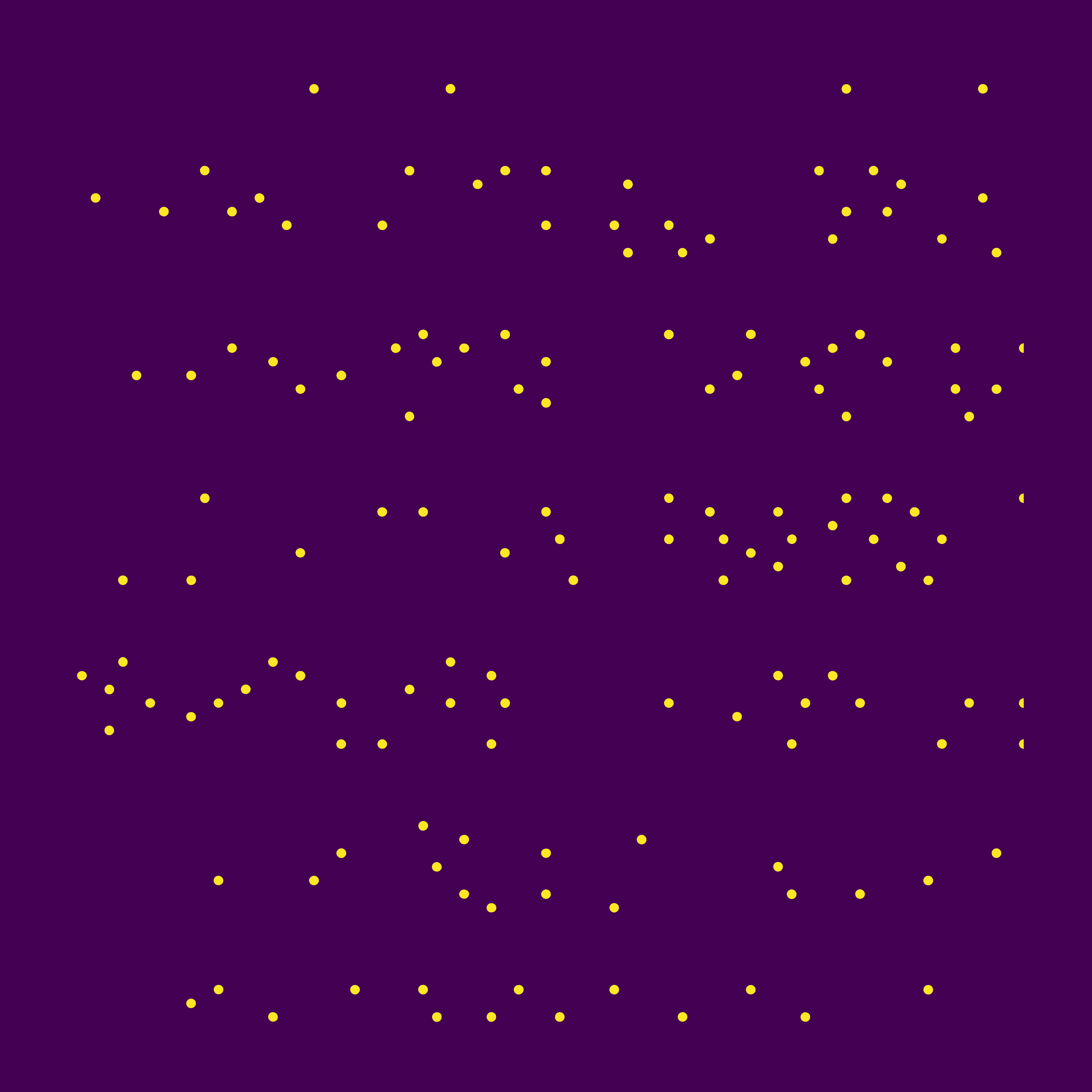}}\\
	\subfloat[Mask]{\includegraphics[width=.24\textwidth,trim={80cm 80cm 80cm 80cm},clip]{mask8k}} \hspace{0.1cm}
	\subfloat[DOINN]{\includegraphics[width=.24\textwidth,trim={80cm 80cm 80cm 80cm},clip]{origin8k}} \hspace{0.1cm}
	\subfloat[DOINN-FS]{\includegraphics[width=.24\textwidth,trim={80cm 80cm 80cm 80cm},clip]{gpp8k}}
	\caption{Visualization of large tile simulation. (a)-(c) are input mask, contour prediction with default DOINN and contour prediction with large tile simulation scheme. (d)-(f) are partial zoom-in view at the same location of (a)-(c) respectively.}
	\label{fig:8k}
\end{figure}
\fi

\begin{table}[tb!]
	\centering
	\caption{Large Tile Simulation Scheme.}
	\label{tab:fs}
	\setlength{\tabcolsep}{13pt}
	\renewcommand{\arraystretch}{.88}
	\begin{tabular}{c|cc}
		\toprule
		ISPD-2019-LT& mPA (\%) & mIOU (\%) \\ \midrule
		DOINN    & 96.30     & 92.03     \\
		DOINN-LT & \textbf{99.25}    & \textbf{98.23}     \\ \bottomrule
	\end{tabular}
\end{table}

\section{Conclusion}
\label{sec:conclu}

In this paper, we tackle the ML-based lithography simulation problem and discuss the drawbacks of state-of-the-art solutions.
We argue that pure convolution-based structures are lacking in efficiency to capture mask semantic information.
To address this problem and build an efficient ML-based lithography simulator, we propose a dual-band optics-inspired neural networks with two major perception paths.
The global perception path is equipped with an optimized Fourier Unit that has the ability to obtain resist intensity information while the local perception path works uses a typical convolutional approach to gather contour details.
Our experiments demonstrate that our framework has the ability to perform lithography simulation more efficiently and effectively compared to previous work. 
Future work includes bringing more accurate physical lithography models with state-of-the-art OPC solutions for training and evaluation,
using more stringent benchmarking criteria for approximating actual silicon rule checks,
and 
incorporating inverse lithography technologies with DOINN for direct mask optimization.

\input{main.bbl}

\renewcommand\thesection{\Alph{section}}
\setcounter{section}{1}
\section*{APPENDIX}
\subsection{The DOINN Architecture}
\subsubsection{GP Layers}
\begin{table}[h]
	\centering
	\caption{GP Layers.}
	\label{tab:gplayers}
	\setlength{\tabcolsep}{10pt}
	\renewcommand{\arraystretch}{1}
	\begin{tabular}{l|l}
		\toprule
		GP Layers   & Output     \\ \midrule
		AvePooling     & 256$\times$256$\times$1  \\ \midrule
		FFT         & 256$\times$129$\times$1  \\ \midrule
		LiftChannel & 256$\times$129$\times$16 \\ \midrule
		MatMul     & 256$\times$129$\times$16 \\ \midrule
		iFFT        & 256$\times$256$\times$16 \\ \bottomrule
	\end{tabular}
\end{table}

\begin{itemize}
	\item AvePooling: 8$\times$8 average pooling with stride 8.
	\item FFT: Fourier Transform is performed channel-wise. Only first 50$\times$50 coefficients are kept, while others are set to zero. This corresponds to the ``truncation'' discussed in \Cref{sec:gp}.
	FFT of a real matrix is Hermitian, therefore only half of the parameters are stored.
	\item LiftChannel: Linear operation in terms of all channels, 256$\times$129$\times$1,1$\times$16 $\rightarrow$256$\times$129$\times$16.
	\item MatMul: Linear transformation on FFT coefficients. Implemented with \texttt{torch.einsum(``bixy,ioxy$\rightarrow$boxy'', input, weights)}.
	\item iFFT: Inverse Fourier Transform performed channel-wise. The output nodes are activated with leakyReLU(0.1).
\end{itemize}

\subsubsection{LP Layers}
\begin{table}[h]
	\centering
	\caption{LP Layers.}
	\label{tab:lplayers}
	\setlength{\tabcolsep}{10pt}
	\renewcommand{\arraystretch}{1}
	\begin{tabular}{l|l|l}
		\toprule
		LP Layers  &Kernel/Stride & Output     \\ \midrule
		conv1    &$4\times4\times1\times4$,2      & 1024$\times$1024$\times$4  \\ \midrule
		vgg1     &$3\times3\times1\times4$,1     & 1024$\times$1024$\times$4  \\ \midrule
		conv2    &$4\times4\times4\times8$,2 & 512$\times$512$\times$8 \\ \midrule
		vgg2     &$3\times3\times4\times8$,1    & 512$\times$512$\times$8 \\ \midrule
		conv3    &$4\times4\times8\times16$,2    & 256$\times$256$\times$16 \\ \midrule
		vgg3     &$3\times3\times8\times16$,1    & 256$\times$256$\times$16 \\ \bottomrule
	\end{tabular}
\end{table}

\begin{itemize}
	\item conv1-3: A single convolution layer.
	\item vgg1-3: Two identical convolution layers for the given spec with BatchNorm and leakyReLU(0.2).
\end{itemize}
\subsubsection{IR Layers}
\begin{table}[h]
	\centering
	\caption{IR Layers.}
	\label{tab:irlayers}
	\setlength{\tabcolsep}{10pt}
	\renewcommand{\arraystretch}{1}
	\begin{tabular}{l|l|l}
		\toprule
		IR Layers  &Kernel/Stride & Output     \\ \midrule
		dconv1    &$4\times4\times32\times16$,2      & 512$\times$512$\times$16  \\ \midrule
		vgg4     &$3\times3\times16\times16$,1     & 512$\times$512$\times$16  \\ \midrule
		dconv2    &$4\times4\times24\times8$,2 & 1024$\times$1024$\times$8 \\ \midrule
		vgg5     &$3\times3\times8\times8$,1    & 1024$\times$1024$\times$8 \\ \midrule
		dconv3    &$4\times4\times12\times4$,2    & 2048$\times$2048$\times$4 \\ \midrule
		vgg6     &$3\times3\times4\times4$,1    & 2048$\times$2048$\times$4 \\ \midrule
		convr1     &$3\times3\times4\times32$,1    & 2048$\times$2048$\times$32 \\ \midrule
		convr2     &$3\times3\times32\times16$,1    & 2048$\times$2048$\times$16 \\ \midrule
		convr3     &$3\times3\times16\times16$,1    & 2048$\times$2048$\times$16 \\ \midrule
		convr4    &$3\times3\times16\times1$,1    & 2048$\times$2048$\times$1 \\ 		\bottomrule
	\end{tabular}
\end{table}
\begin{itemize}
	\item dconv1-3: Single layer of transposed convolution.
	\item convr1-3: Single convolution layer with ReLU activation.
	\item convr4: Single convolution layer with Tanh activation.
	\item Input of dconv1 is the concatenated feature maps of iFFT and vgg3.
	\item Input of dconv2 is the concatenated feature maps of vgg4 and vgg2.
	\item Input of dconv3 is the concatenated feature maps of vgg5 and vgg1.
\end{itemize}

\subsection{Configurations}
For all the designs, the DOINN is trained with the following configurations.
Environment: NVIDIA A100 GPU, Ubuntu 20.04 LTS, Pytorch 1.9.0
\begin{table}[h]
	\centering
	\caption{Training Configurations.}
	\label{tab:train}
	\setlength{\tabcolsep}{10pt}
	\renewcommand{\arraystretch}{1}
	\begin{tabular}{l|l}
		\toprule
		Max Epoch  &10    \\ \midrule
		Initial Learning Rate    &0.002  \\ \midrule
		Learning Rate Decay Policy     &Step, Every 2 epochs   \\ \midrule
		Learning Rate Decay Factor    &0.5 \\ \midrule
		Batch Size     &16   \\ \midrule
		Optimizer    &Adam\\ \midrule
		Weight Decay     &0.0001 \\ \midrule
		Loss     &MSE \\ 	\bottomrule
	\end{tabular}
\end{table}

\end{document}

%% file: acm-setting.tex
\usepackage{lipsum}
\usepackage{titlesec}

\usepackage{graphicx}
\usepackage{amsmath}
\usepackage{amssymb}
\usepackage{mathtools}
\usepackage{comment}
\usepackage[subrefformat=parens,labelformat=parens]{subfig}
\usepackage{bm}
\usepackage{multirow}
\usepackage{threeparttable,booktabs}
\usepackage{blkarray}
\usepackage{tikz}
\usepackage{balance}
\usepackage{courier}                                       
\usepackage{cleveref}                                      
\usepackage[mathcal]{eucal}
\usepackage[noend]{algpseudocode}
\algrenewcommand\textproc{\texttt}
\makeatletter\let\float@addtolists\relax\makeatother
\usepackage{algorithm}

\DeclarePairedDelimiter\floor{\lfloor}{\rfloor}
\usepackage{filecontents}                                  
\usepackage{pgfplots}
\usepackage{pgfplotstable}
\pgfplotsset{compat=newest}
\usepackage[figuresright]{rotating}

\renewcommand{\vec}[1]{\boldsymbol{#1}}

\newtheorem{mydefinition}{\textbf{Definition}}
\newtheorem{myproblem}{\textbf{Problem}}

%% file: time.tex
\begin{tikzpicture}
    \pgfplotsset{
        width =0.6\textwidth,
        height=0.4\textwidth,
        every axis plot/.append style = {font = \small},
        /pgfplots/bar cycle list/.style={/pgfplots/cycle list={%
            {NVblack,fill=NVgreen},
            }
            },
        }
        \begin{axis}[
                ybar=2.5pt,
                enlarge x limits=0.2,
                bar width=10pt,
                legend style={
                    draw=none,
                    at={(0.5,1.0)},
                    anchor=south,
                    legend columns=-1
                },
                area legend,
                ylabel={$\mu m^2$ / s},
                symbolic x coords={UNet,DAMO,Ours,Ref},
                xtick=data,
                ymin=0,
                ytick={0,15,30,45},
                ylabel near ticks,
                x tick label style={rotate=0},
                nodes near coords,
            ]
            \addplot  coordinates {(UNet,41) (DAMO,4.76) (Ours,34)(Ref,0.4)};

        \end{axis}
\end{tikzpicture}

%% file: 24opc.tex
\pgfplotsset{
	width =0.6\textwidth,
	height=0.4\textwidth
}
\begin{tikzpicture}[scale=1]
	\begin{axis}[minor tick num=0,
		xmax=24,xmin=1,
		ymax=1.0, ymin=0.8,
		yticklabel style={/pgf/number format/.cd, fixed, fixed zerofill, precision=2, /tikz/.cd},
		xlabel={OPC Iterations},
		ylabel={mIOU},
		xlabel near ticks,
		legend style={
			draw=none,
			at={(0.50,1.2)},
			anchor=north,
			legend columns=3,
		}
		]
		\addplot +[line width=0.7pt] [color=NVgreen,   solid, mark=o]  table [x={iter},  y={doinn}]  {24opc.dat};
		\addplot +[line width=0.7pt] [color=NVmgrey,  solid, mark=diamond]   table [x={iter},  y={unet}]  {24opc.dat};
		
		\legend{DOINN, UNet}
		
	\end{axis}
\end{tikzpicture}